\documentclass[12pt]{spieman}  
\usepackage{amsmath,amsfonts,amssymb}
\usepackage[final]{graphicx}
\usepackage{setspace}
\usepackage{tocloft}
\usepackage{lineno}
\usepackage{booktabs}
\usepackage{array}
\title{A High-Bandwidth Backplane for Wideband Radio Interferometers and Integration with the CHORD Telescope Correlators}

\author[a,b,c*]{Wellington Avelino}
\author[a,b,c]{Joshua Montgomery}
\author[a]{Jean-Francois Cliche}
\author[b]{Graeme Smecher}
\author[a,b,c]{Matt Dobbs}
\affil[a]{McGill University, Physiscs Department, 3600 University Street, Montréal, Québec, Canada}
\affil[b]{t0.technology, 300 rue Léo-Pariseau, Montréal, Québec, Canada}
\affil[c]{Trottier Space Institute, 3550 University Street, Montréal, Québec, Canada}

\cftpagenumbersoff{figure}
\cftpagenumbersoff{table} 
\begin{document} 
\maketitle

\begin{abstract}
Wide-band correlators for radio astronomy interferometers demand accurate, scalable signal processing backends that sustain high aggregate throughput while preserving stable timing alignment across a large number of signal chains. Addressing these coupled constraints is challenging without a co-designed backplane architecture. We present a backplane system developed for the Control Readout System (CRS) Field Programmable Gate Array (FPGA) platform, extending its capabilities to support high channel count wideband radio correlators. The design and validation process are guided by the requirements of the Canadian Hydrogen Observatory and Radio-transient Detector (CHORD) telescope, which provides a representative use case for the system’s performance evaluation. Each backplane hosts up to four CRS boards, and a single crate integrates four interconnected backplanes to accommodate sixteen CRS modules (for a total of 128 digitized inputs per crate) interconnected through 25 Gb/s per-lane links and an on-backplane data shuffle network. Comprehensive validation demonstrates robust high-speed link performance, sub-sampling-level timing alignment stability, and stable thermal and power behavior across environmental variations. The architecture delivers observatory-scale scalability, providing a low-jitter and phase-consistent foundation for wideband digital correlators. By supporting expansion from single-crate to multi-crate configurations without firmware modification, the system offers a practical and flexible path toward next-generation interferometers requiring tightly synchronized, high-throughput digital backends. 
\end{abstract}


{\noindent \footnotesize\textbf{*}Wellington Avelino,  \linkable{wellington.deoliveiraavelino@mcgill.ca} }

\vspace{\baselineskip}
{\noindent\small\textbf{Preprint notice.} This manuscript is a preprint version and has not yet published. It has been prepared for possible submission to \textit{Journal of Astronomical Telescopes, Instruments, and Systems - SPIE.}}

\begin{spacing}{2}   

\section{Introduction}
\label{sect:intro}  
Advances in commercial digital signal processing technology have shifted the focus in radio astronomy from traditional single-dish telescopes to large interferometric arrays such as SKA\cite{Carilli2004}, LOFAR\cite{deVos2009}, MWA\cite{Lonsdale2009}, LWA\cite{Ellingson2009}, UTMOST\cite{Caleb2016}, CHIME\cite{Bandura2014}, HERA\cite{DeBoer2014}, and HIRAX\cite{Newburgh2016}. These instruments employ hundreds to thousands of smaller antenna elements distributed over wide areas, achieving high angular resolution, large collecting area, and large fields of view, by digitally combining the raw data from a large number of interferometric baselines. Compared to monolithic designs, distributed architectures provide denser visibility sampling and sustain high sensitivity across broad frequency ranges. For example, CHIME employs this technology to detect fast radio bursts\cite{Amiri2025}, map diffuse emission\cite{Amiri2024}, and tracing the large-scale structure of the universe through 21-cm intensity mapping\cite{Amiri2023}. The real-time digital data processing and full mesh networking necessary to correlate data from these arrays sets stringent requirements on the phase synchronization of the distributed digitization and signal processing systems, as well as reliable high-bandwidth data transport between processing elements when required by the chosen backend partitioning. The advanced backplane architecture described in this work enables the modular and distributed digitization and signal-processing system to operate cohesively, achieving high precision measurements across large interferometric arrays.

\noindent Multi-antenna radio interferometers perform digital spatial-correlation (X-engine) between all antenna feeds over a broad bandwidth, resulting in substantial high-bandwidth data exchange between receiver processing elements. Signals captured from antennas undergo digitization and frequency channelization—often referred to as the D-engine and F-engine, respectively—before being routed through the backplane to the X-engine. For instance, CHIME handles over 6.5 Tb/s of internal data traffic\cite{Bandura2014}. For CHORD, despite a reduced number of antenna signals (1024 vs. 2048 in CHIME), the system bandwidth almost triples (from 400 MHz to 1200 MHz), increasing the total data throughput to approximately 10 Tb/s\cite{Vanderlinde2019}, which can be carried by different interconnect strategies; the backplane links described here provide an intra-crate option alongside switch-based transport.

\noindent The backplane provides not only a physical pathway for data movement and a mechanical carrier for multiple processing boards, but also the timing and coordination backbone that preserves coherence across the entire array\cite{Bandura2016}. Backplane boards serve as a key enabler of system scalability, providing power distribution, timing/synchronization high-speed interconnections that aggregate and distribute data among DFX-engine modules embedded within the signal-processing boards. They implement a fixed-topology data-exchange fabric required for the intra-crate “corner-turn” operation, in systems that implement this stage in hardware, wherein each F-engine circuit board shares frequency-partitioned streams from all antennas, before routing the aggregated outputs to centralized GPU-based X-engine units for spatial correlation and image construction. As the number of active antenna feeds grows, the aggregate data throughput increases proportionally, demanding a concurrent communication infrastructure capable of moving tens of Tb/s between boards with stable timing alignment. This degree of parallelization requires all modules to operate in precise synchrony—any deviation in latency or clock phase can decorrelate signals and degrade coherence across the field of view.

\noindent In large interferometric arrays, maintaining phase coherence across hundreds of distributed receivers is essential for preserving sensitivity and imaging fidelity. For instance, clock jitter and skew have been analyzed in interferometric radiometers, demonstrating that residual sampling timing errors increase visibility uncertainty and degrade system SNR\cite{Zhang2014}. Similarly, sub-nanosecond channel delay mismatches, e.g. 10–100 ps, lead to measurable gain loss and pointing errors of a few percent of the beam width in phased arrays\cite{Schediwy2010} particularly when such delays are time-varying or not fully calibrated. More generally, array-theory analyses show that uncorrelated phase errors across elements reduce the effective beamforming gain by a factor of exp($-\sigma_\phi^2/2$), where $-\sigma_\phi$ is the RMS phase error\cite{Johnson1992}. Maintaining sub-nanosecond timing alignment across distributed acquisition channels represents a significant systems-engineering challenge. At the same time, transporting the resulting multi-gigabit data rates between processing modules demands scalable, high-bandwidth interconnects. Although commercial interconnects support high-speed data transfer, their cost, power, and integration complexity become prohibitive at the scales required. A customized backplane architecture therefore becomes not merely an engineering convenience but a scientific necessity.

\noindent This work introduces such a backplane system, developed for wideband radio correlators. Built around t0.technology's Control Readout System digital-signal-processing platforms \cite{Montgomery2024} and informed by lessons learned from the CHIME correlator, the design implements high density all-to-all interconnect capable of timing accuracy and data rates necessary for next-generation instruments such as the CHORD telescope.

\section{Evolution and Requirements Leading to Custom Backplane Systems}
Early radio astronomy backplanes in the 1980s and 1990s employed basic parallel architectures, typically using VMEbus \textit{(VERSAmodule Eurocard)} standards, capable of 16-bit modular connectivity and typically limited to around 40 Mbps via single-ended signaling, the backplane links described here provide an intra-crate option alongside switch-based transport depending on the bus variant\cite{Heath1989}. VME lacked precise synchronization, introducing timing variability that limited large-scale interferometric applications. Despite technological constraints, these initial architectures established the fundamental modular approach essential for later instrumentation advances.

\noindent In the 2000s,\textit{Compact Peripheral Component Interconnect} (CompactPCI, cPCI) emerged as an important off-the-shelf backplane standard for industrial, medical, aerospace, and telecommunications applications\cite{McMahon2008}. Derived from the PCI bus but implemented in a Eurocard format, cPCI supported 32- or 64-bit parallel transfers at 33–66 MHz, achieving peak data rates of up to 528 MB/s\cite{KleinesCompactPCI}. In radio astronomy cPCI still relied on dedicated external clock and timing distribution hardware, rather than providing precise synchronization inherently at the backplane level. Thus, while bandwidth increased, stable timing remained external to the backplane fabric.

\noindent In the 2010s, the emergence of \textit{Advanced Telecommunications Computing Architecture} (ATCA) and PCIe-based backplanes marked a transition toward switched-fabric, high-throughput modular systems. These platforms were adopted in large-scale interferometric arrays and very-long-baseline instruments such as the Event Horizon Telescope (EHT)\cite{Doeleman2023}. These commercial fabrics improved aggregate throughput but still required careful integration to meet the phase-stability demands of precision correlators.

\noindent By the mid-2010s, custom FPGA-based backplane architectures became common for large-scale radio interferometers. A prominent example is the CHIME correlator\cite{Bandura2016}, which uses custom ICE backplanes to implement terabit-per-second corner-turn networks. Similar backplanes have been deployed in other instruments, including LOFAR upgrade systems, MeerKAT digital backends, and SKA pathfinder projects. These efforts demonstrated that purpose-built backplanes could achieve the stable latency and synchronization performance required for wideband correlators.

\noindent Since the 2020s, optical interconnects have gained increasing relevance in radio astronomy, driven by the extreme data-rate and scalability demands of next-generation instruments such as the Square Kilometre Array (SKA). Studies and prototype implementations addressing SKA-class correlator and beamforming systems have explored the use of optical interconnections, including backplane-level optical fabrics, to support the required data redistribution and processing\cite{Hampson2015}. Unlike electrical backplanes, optical fabrics provide order-of-magnitude higher aggregate bandwidth, reduced signal degradation over distance, and reduced susceptibility to crosstalk and electromagnetic interference. Early adoption in SKA pathfinder projects demonstrates their potential for low-latency, petabit-scale data transport, although they introduce higher integration complexity, increased power budgets, and non-trivial clock-recovery\cite{McCool2011}.

\noindent Table \ref{tab:backplane_evolution} summarizes the key metrics and performance characteristics of prominent architectures used over the past four decades.

\begin{table}[H]
\caption{Evolution of backplane technologies, summarizing key metrics such as achievable data rates, signaling methods, synchronization precision, scalability, deployment cost, and representative telescope projects.}
\label{tab:backplane_evolution}
\centering
\footnotesize
\renewcommand{\arraystretch}{1.4}
\setlength{\extrarowheight}{2pt}
\begin{tabular}{
>{\centering\arraybackslash}m{1.6cm}
>{\centering\arraybackslash}m{1.6cm}
>{\centering\arraybackslash}m{1.4cm}
>{\centering\arraybackslash}m{1.8cm}
>{\centering\arraybackslash}m{1.7cm}
>{\centering\arraybackslash}m{1.6cm}
>{\centering\arraybackslash}m{1.4cm}
>{\centering\arraybackslash}m{2.2cm}
}
\toprule
Era & Technology & Data Rate per link & Signaling & Sync. Precision & Modularity/ Scalability & Qualitative Cost & Example Telescope \\
\midrule
1980--1990 & VME & $\sim$40 Mbps & Parallel, Single-ended & Low & Moderate & Low & Early VLA-era arrays \\
2000s & CompactPCI & $\sim$100 Mbps & Differential, Serial & Moderate & Moderate/ High & Moderate & VLA upgrade, Early LOFAR \\
2010s & ATCA, PCIe & 1--40 Gbps & Differential, Serial & High & High & Moderate & ALMA, EHT \\
Mid-2010s & Custom backplane & 10--100 Gbps & High-speed serial & Very High & Very High & Low/    Moderate & CHIME, HIRAX \\
2020s+ & Optical backplane & $>$100 Gbps (Tbps+) & Optical & Extremely High & Extremely High & Very High & Next-gen telescopes \\
\bottomrule
\end{tabular}
\end{table}

\noindent As summarized in Table \ref{tab:backplane_evolution}, successive generations of backplane technologies have delivered steady increases in per-link data rate and modular scalability. At the same time, systems operating in regimes that demand very high timing precision exhibit a growing emphasis on controlled latency and phase stability. In these regimes, calibration effort and system-level integration complexity tend to scale more rapidly than raw link bandwidth, indicating a shift in the architectural constraints as instruments grow in size and bandwidth.

\noindent This evolution reflects a consistent trend in the evolution of digital backends for radio interferometry: as instruments scale, interconnects must be co-designed with the timing, calibration and signal integrity constraints of the telescope, not just peak bandwidth. New commercial link generations increase nominal data rates, but operating near their limits adds integration complexity without necessarily improving timing stability or calibration effort. The architecture presented here therefore prioritizes instrument-driven performance—stable timing alignment, controlled latency behavior, and manageable system complexity——over adopting the highest available commercial data rates.

\section{Backplane Design for Timing-Stable and Scalable Channelization}
\begin{figure}[h]
\begin{center}
\includegraphics[width=\linewidth, trim=35mm 5mm 5mm 37mm, clip]{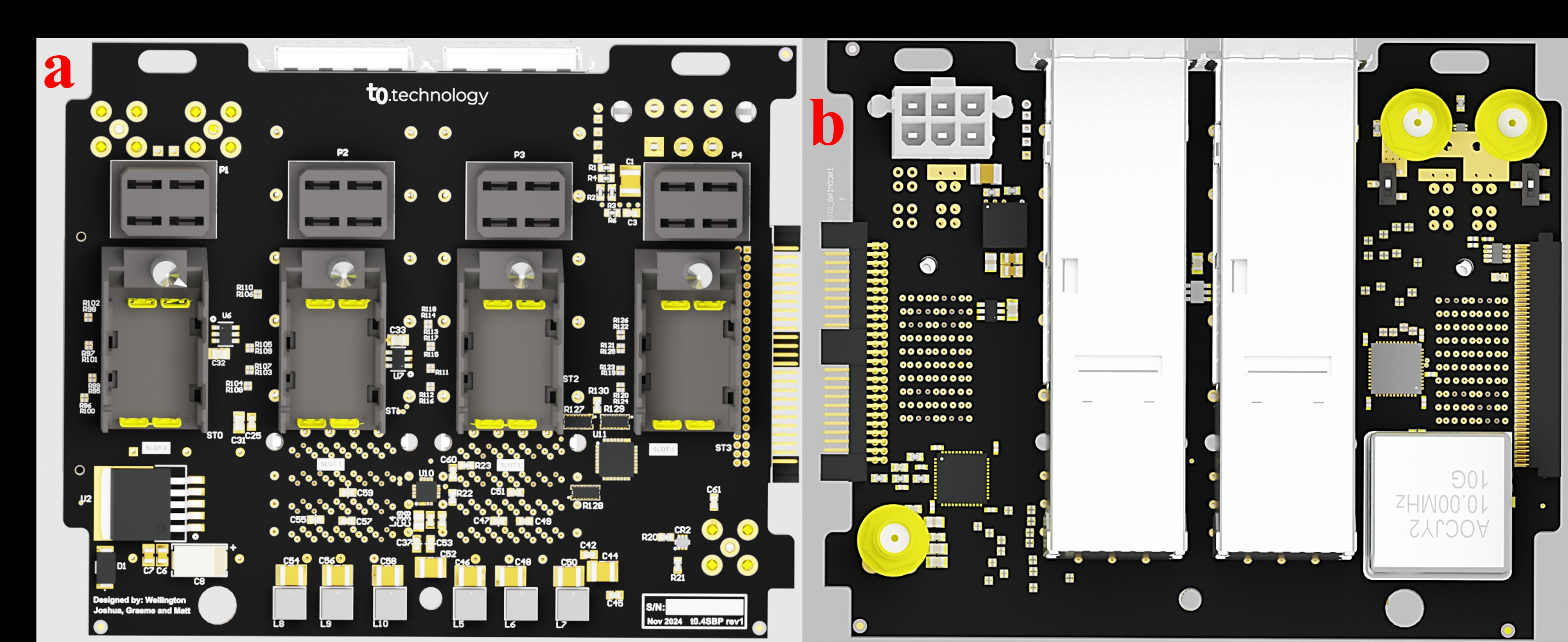}
\end{center}
\caption{ Customized backplane board designed for modular CRS board integration: a) Front-side view showing connectors for up to four Control Readout System (CRS) boards; b) Back-side view detailing connection for expanding: QSFP data interfaces and power distribution connectors.}
\label{aba:fig1}
\end{figure}

The backplane system developed in this work adopts a requirement-driven architecture that prioritizes stable timing, high signal integrity, high-speed interconnectivity, and cost-balanced scalability for CRS-based correlator systems. Each backplane module interconnects up to four CRS boards through passive high-speed channels operating at 25 Gbps per lane. The board views are illustrated in Fig.\ref{aba:fig1}.

\noindent Each backplane board integrates four high-speed zQSFP+ connectors that distribute data among four other backplane boards within a 6U, 19-inch crate configuration (128 inputs per crate). The CHORD array, designed for 512 digital inputs, will therefore employ four such crates to achieve full operational capacity. 

\noindent Solder-free connectors were selected to reduce impedance discontinuities and insertion loss for high speed interconnects on the backplane —preserving signal integrity at multi-gigabit speeds\cite{Mattsson2014}. The backplane PCB stack-up uses a cost-balanced low-loss substrate (Panasonic MEGTRON-4) with passive differential stripline routing, enabling a compact 14-layer design within an 85 × 104 mm footprint while maintaining controlled impedance and low crosstalk. Backdrilling was employed on multi-gigabit signal vias to minimize stub-induced reflections and further improve signal integrity. Together, these channel characteristics, combined with standard adaptive equalization available in contemporary FPGA transceivers, support stable 25 Gb/s intra-backplane links between CRS boards. Inter-backplane connectivity is implemented through quad small form-factor pluggable 28 (QSFP28) interfaces operating at comparable per-lane data rates, extending the signal path while maintaining stable link margins. 

\begin{figure}[H]
\begin{center}
\includegraphics[width=16cm]{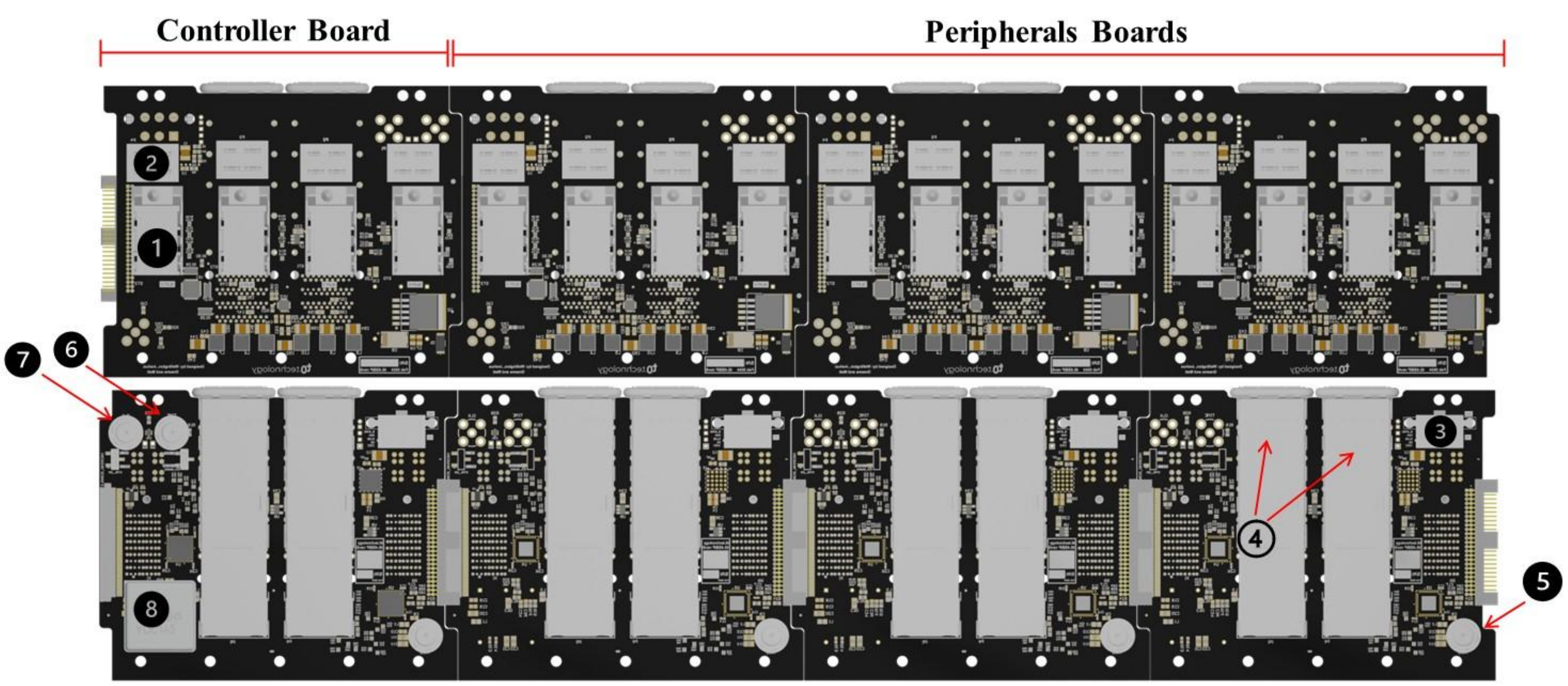}%
\end{center}
\caption{ Scalable configuration of interconnected backplane boards: Physical layout showing: (1) data and (2) power connector for CRS boards; (3) 12 VDC power distributed from the subrack; (4) double-height zQSFP+ (QSFP28-compatible) cages; (5) user-defined, (6) IRIG-B and (7) bi-directional clock SMA connector; (8) oven-controlled crystal oscillator (OCXO).}
\label{aba:fig2}
\end{figure}

\noindent Within each module, the backplane distributes power and reference signals and provides fixed-topology high-speed data exchange among the CRS boards. System scalability is achieved by cascading multiple backplane modules. A designated controller backplane distributes timing and configuration control to peripheral backplanes via side-edge connectors, maintaining timing alignment across the stack, see Fig.\ref{aba:fig2}. High-speed data exchange between backplanes is implemented using passive QSFP28 links, enabling expansion without active switching. In this configuration, four interconnected backplanes support 16 CRS boards, corresponding to a 128-input correlator system.

\begin{figure}[H]
\begin{center}
\includegraphics[width=16cm]{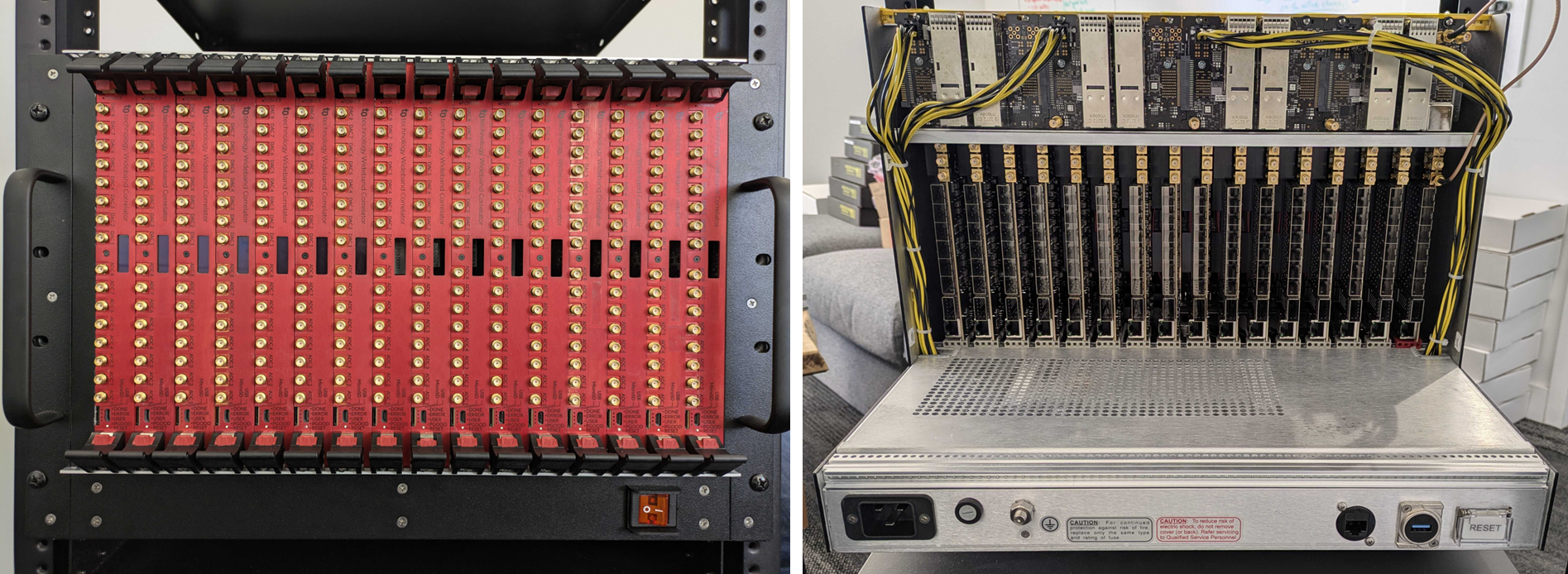}%
\end{center}
\caption{ Example of a fully assembled crate accommodating 16 interconnected CRS boards.}
\label{fig:backplane3}
\end{figure}

\noindent System synchronization is achieved by distributing a 10 MHz reference and IRIG-B timing signals from a controller board via low-voltage differential signaling (LVDS). A low-additive-jitter fan-out buffer delivers the reference directly to all slots in parallel, minimizing phase-noise accumulation and avoiding cascaded clock regeneration. When an external reference is not available, the backplane operates in a self-contained mode: an onboard OCXO provides the master 10 MHz signal (Fig.\ref{aba:fig2}-8), and the CRS boards synthesize and distribute IRIG-B, enabling standalone deployments without dependence on external timing instruments. 
The design choices outlined here provide the foundation for the high-speed performance and timing stability evaluated in the following section, where bit error rates (BER) measurements and eye-diagram analyses quantify the backplane’s signal integrity and synchronization performance.

\section{Backplane System Performance} 
In this section, the experimental assessment of the backplane is reported with emphasis on the readiness for the CHORD telescope under full-crate operating conditions. The evaluation focused on the following key performance indicators (KPIs): (i) multi-gigabit data-link signal integrity and bit-error-rate (BER) performance, (ii) clock distribution quality and jitter performance, (iii) power delivery stability under full-load conditions, and (iv) thermal behavior within a fully populated crate. The following subsections present the validation results for each of these performance domains under representative operational scenarios. 

 \subsection{Backplane Interconnect Reliability: Data-link signal integrity and BER performance}
CHORD must sustain an aggregate data throughput of up to 10 Tb/s across 1024 feeds, requiring continuous, error-free inter-board communication under worst-case data traffic conditions. While the baseline CHORD architecture performs crate-level aggregation through a centralized 100-GbE switching layer, the passive backplane mesh described in the previous section provides a fixed-topology, infrastructure-minimizing option for intra- and inter-crate data exchange. Validation of these passive high-speed channels therefore serves both as performance qualification and architectural risk mitigation.

\noindent To qualify link robustness, all on-board and backplane high-speed lanes were validated at 25 Gb/s per lane using the integrated bit error rate tester (IBERT) framework. This data rate approaches the device-rated line speed of the Zynq RFSoC UltraScale+ transceivers, providing realistic stress conditions while preserving margin protocol overhead. Tests were conducted under simultaneous bidirectional loading across all populated slots to emulate worst-case concurrent operation and evaluate potential channel-to-channel crosstalk.

\begin{figure}[H]
\begin{center}
\includegraphics[width=16cm]{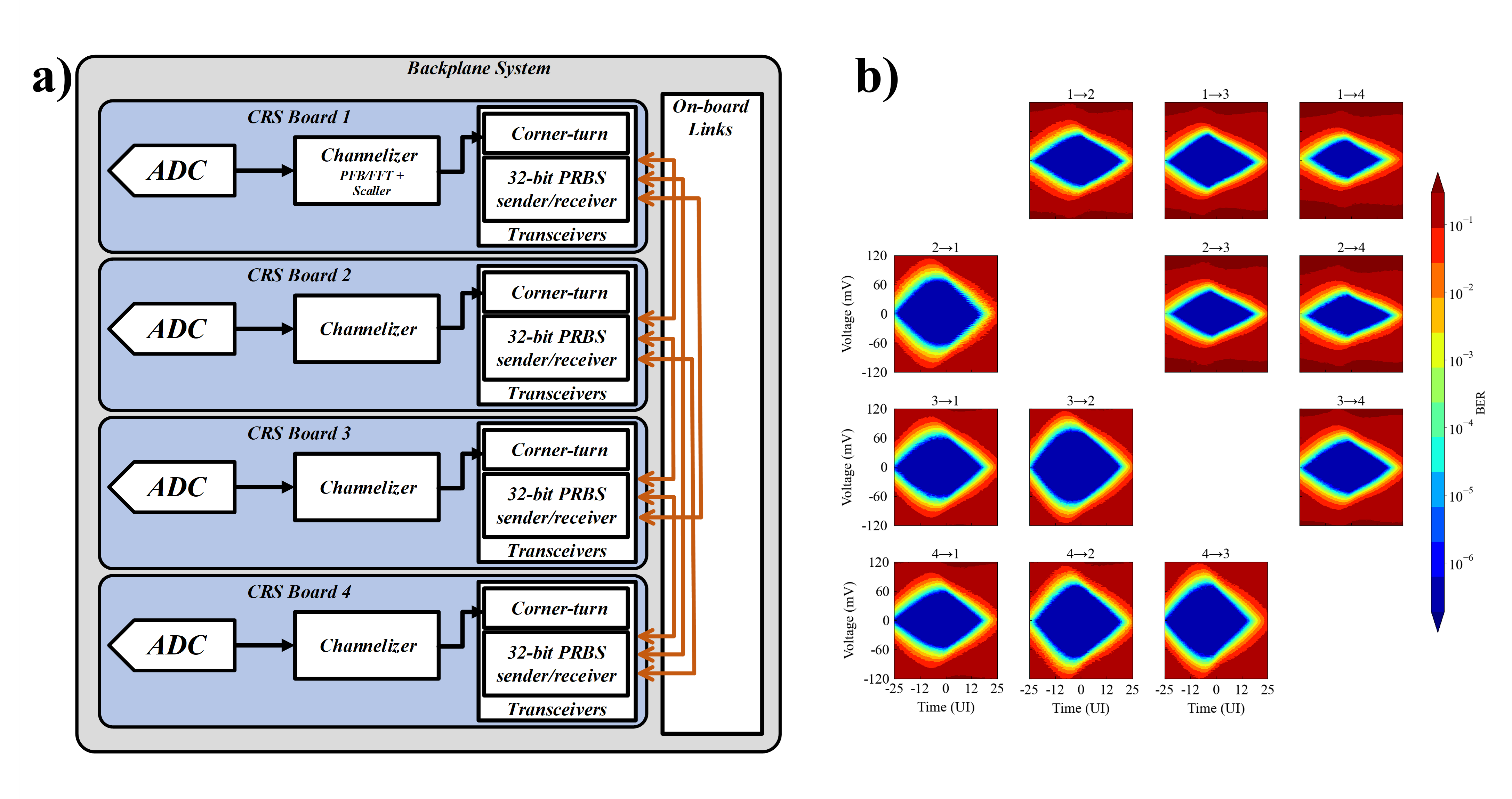}%
\end{center}
\caption{ (a) Simplified diagram for the backplane on-board links test setup: CRS boards are interconnected through the backplane, the IBERT core is instantiated, driving a Pseudo-random Binary Sequence (PRBS) generator/checker (32-bit pattern). (b) BER-based eye-diagram matrix for the links under test, captured with all links transmitting and receiving simultaneously. Across all lanes, the eyes remain open and no bit errors were observed over the acquisition window, indicating error-free operation.}\label{fig:backplane4}
\end{figure}

\noindent Fig.\ref{fig:backplane4}-a presents the validation results for slot-to-slot links within a fully populated backplane board. BER-based statistical eye scans demonstrate clear timing and amplitude margins across all lanes, Fig.\ref{fig:backplane4}-b. No bit errors were observed over more than 10 Tbits transmitted per channel. Selective activation of adjacent lanes produced no measurable degradation in eye opening or BER performance, confirming immunity to inter-channel crosstalk under simultaneous operation.

\begin{figure}[h]
\begin{center}
\includegraphics[width=16cm]{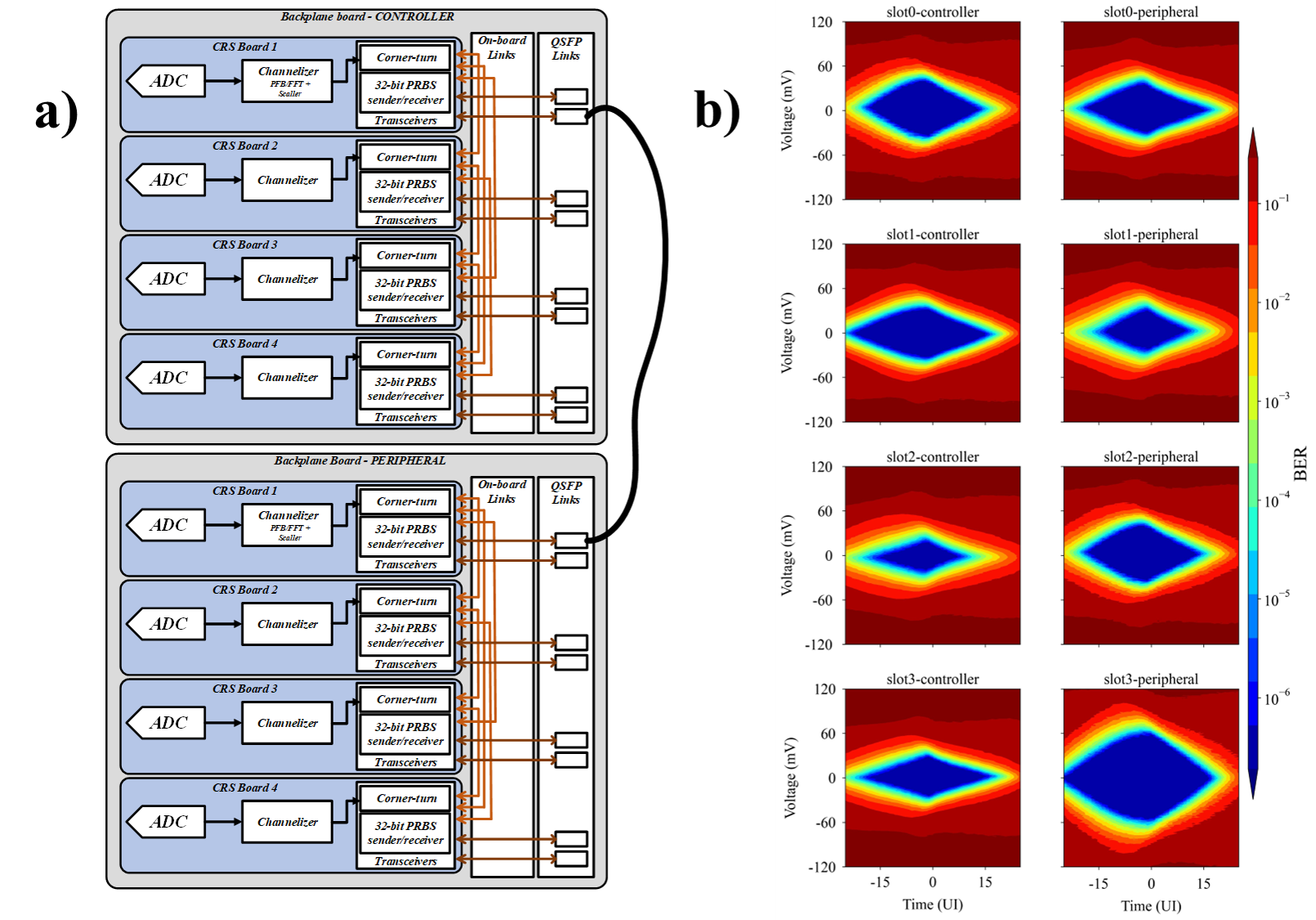}%
\end{center}
\caption{ (a) Simplified schematic of the inter-board configuration showing data routing between two fully populated backplane boards through slot-aligned QSFP28 connections (S0–S3). With this slot-to-slot wiring, only the lanes associated with the four aligned QSFP28 positions are exercised, resulting in eight active 25-Gb/s lanes (four per direction) (b) BER-based eye-diagram measurements for each active lane at nominal operating conditions.}
\label{fig:backplan5}
\end{figure}

\noindent Inter-crate links implemented through equal-length QSFP28 interfaces were validated under identical 25 Gb/s conditions. Figures \ref{fig:backplan5}(a-b) present a simplified diagram and the eye diagram for 25 Gbps data transmission respectively. Both PCB backplane channels and QSFP28 cable paths satisfy the target BER requirements. The shorter impedance-controlled stripline traces within the passive backplane exhibit larger vertical and horizontal eye margins due to reduced insertion loss and fewer discontinuities. The QSFP28 paths introduce additional connector transitions and longer electrical lengths, resulting in reduced—but fully compliant—timing and voltage margins.

\noindent Collectively, these results confirm that the passive backplane interconnect supports sustained error-free 25 Gb/s operation across all slots under full concurrent loading, satisfying the data integrity requirements for CHORD-class correlator deployments.

 \subsection{Backplane Timing Accuracy: Clock Distribution Quality and Jitter Performance }
\begin{figure}[H]
\begin{center}
\includegraphics[width=16cm]{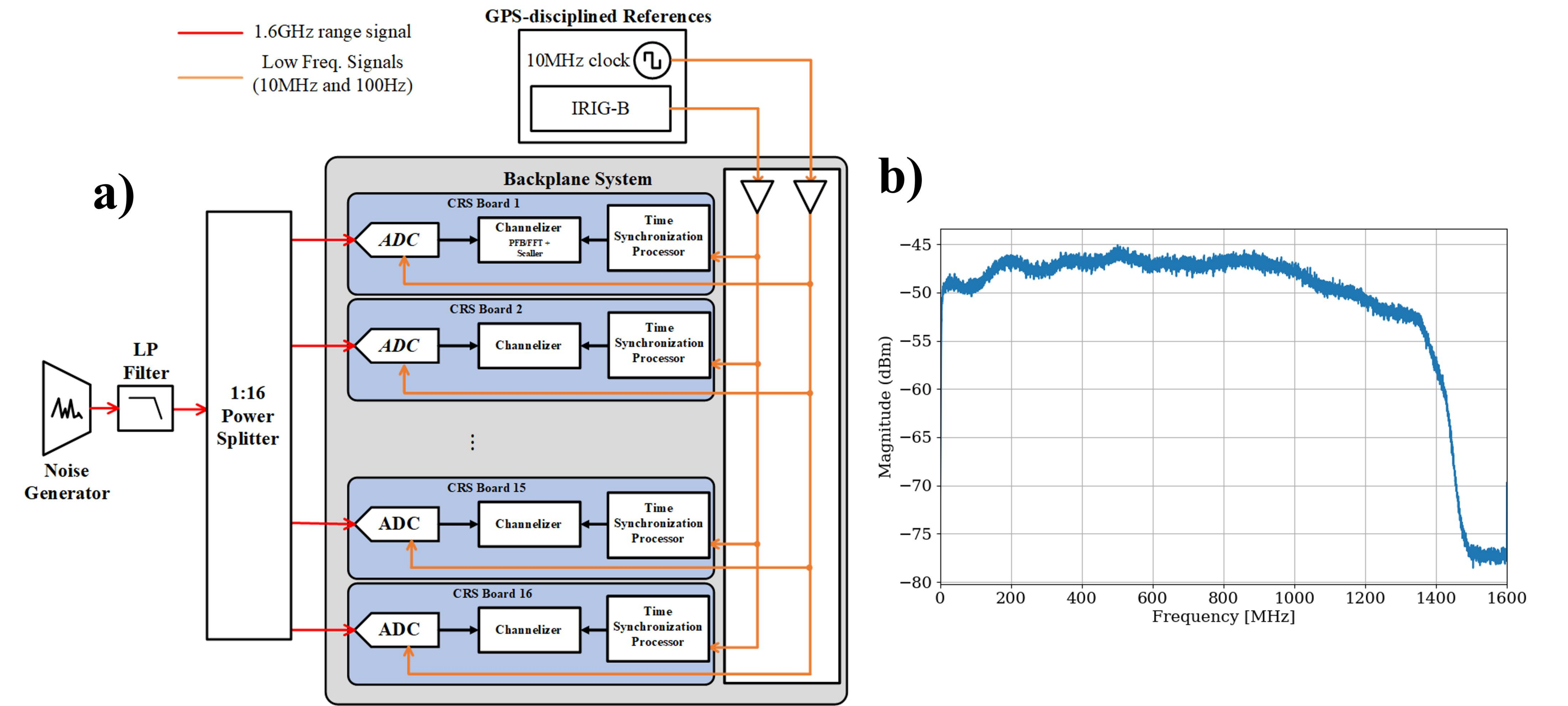}%
\end{center}
\caption{ Experimental setup for synchronization analysis of CRS boards inside the backplane crate. a) Simplified block diagram of the test setup. A noise generator feeds a low-pass filter with 1.4 GHz cutoff, followed by a power divider that distributes the signal to the backplane system with 16 CRS boards. All boards are synchronized by a GPS-disciplined reference providing a 10 MHz clock and IRIG-B timing. b) Frequency response of the injected signal used for the tests.}
\label{fig:backplan6}
\end{figure}

The timing performance of the backplane system is evaluated in the context of interferometric requirements in this section. By driving multiple CRS boards with a shared reference signal, we extract inter-board delay, jitter, and drift metrics that quantify both instantaneous and long-term temporal stability.
A broadband noise source and a 1.4 GHz lowpass filter are connected to a 1:N power splitter, which distributes the same signal simultaneously to one input of each board mounted on the backplane, as illustrated in  Fig.\ref{fig:backplan6}-a. This guarantees that every board receives an identical reference waveform, so any measured timing differences arise from variations in the clock signal paths, connectors, or board circuitry. The use of a broadband noise source is intentional: its wide spectral content excites the full measurement bandwidth of each channel, enabling reliable cross-correlation–based delay estimation.

\noindent Relative time delays between boards are obtained from the generalized cross-correlation (GCC) with Phase Transform (PHAT) function, defined as:

\begin{equation}
    R_{ij}(\tau)=\mathcal{F}^{-1}\!\left\{\,W(f)\,X_i(f)\,X_j^{*}(f)\,\right\}.
\label{aba:eq1}
\end{equation}

\noindent Where $X_i(f)$ and $X_j^{*}(f)$ are the Fourier transforms and its conjugate of the captured signals, $W(f)$ is a frequency-domain weighting function, and $\tau$ is the relative delay. In the PHAT formulation \cite{Knapp1976}, $W(f)=1/|X_i(f) X_j^{*}(f)|$, emphasizing phase information and suppressing amplitude variations. To achieve sub-sample timing precision, we refine the GCC-PHAT peak by zero-padding and a local quadratic (parabolic) interpolation around the maximum, a standard approach in cross-correlation time-delay estimation, yielding sub-sample delay estimates, corresponding to tens of picoseconds at the adopted sampling rate.

\begin{figure}[H]
\begin{center}
\includegraphics[width=16cm]{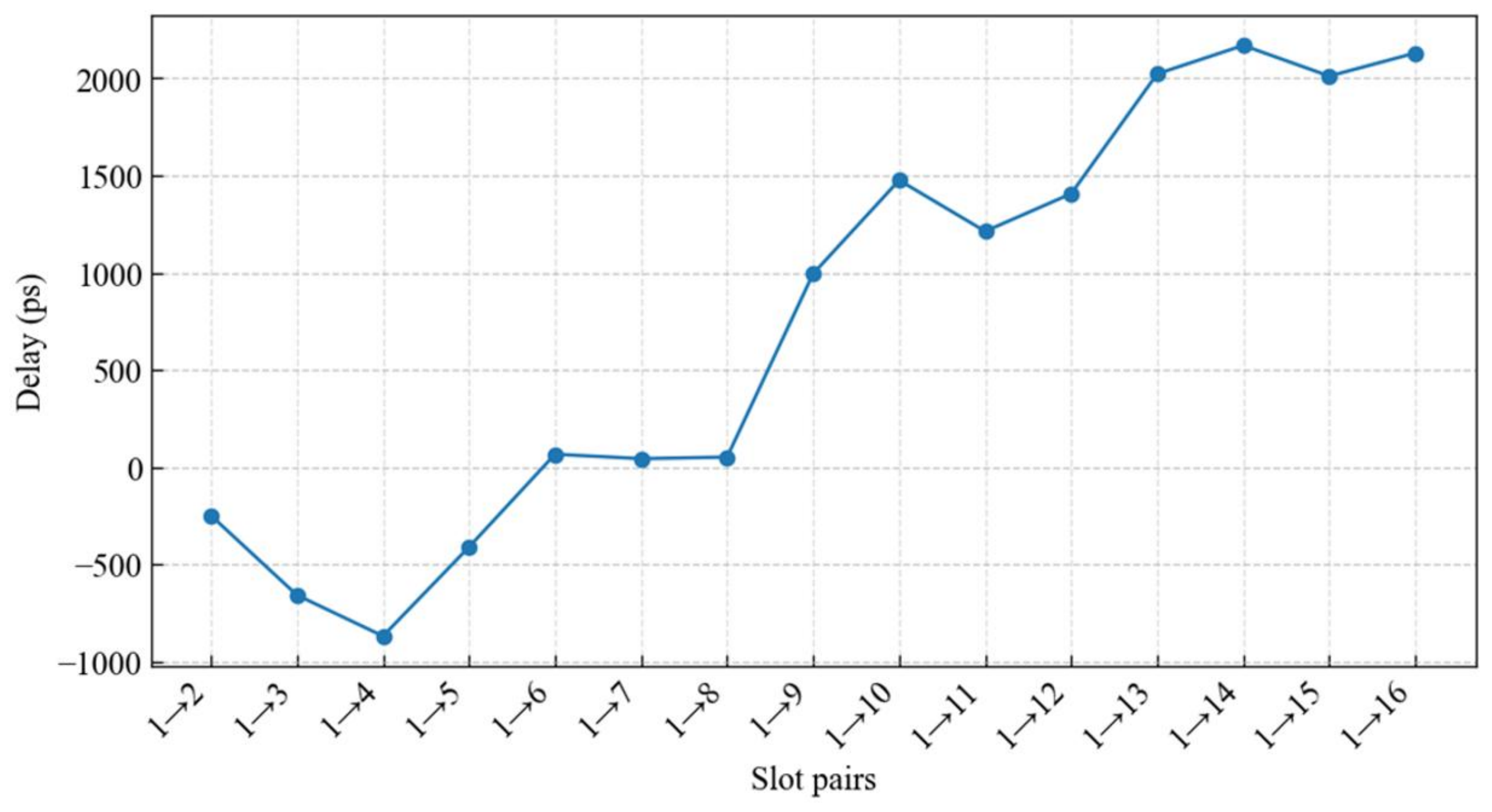}%
\end{center}
\caption{ Relative delay (or acquisition skew) between backplane slot pairs. The x-axis denotes slot-pair indices (1→N), while the left y-axes report the measured delays in picoseconds.}
\label{fig:backplan7}
\end{figure}

Fig. \ref{fig:backplan7} shows the relative delays between backplane slots pairs (slot 1 used as reference).

\noindent The relative delay between boards is not a primary performance metric. Instead, the key requirement is that this delay be repeatable and stable over time. Any fixed latency introduced by the backplane or acquisition chain can be removed through calibration procedures. System performance therefore depends on delay stability and repeatability rather than on the absolute value of the acquisition delay.

\begin{figure}[H]
\begin{center}
\includegraphics[width=16cm]{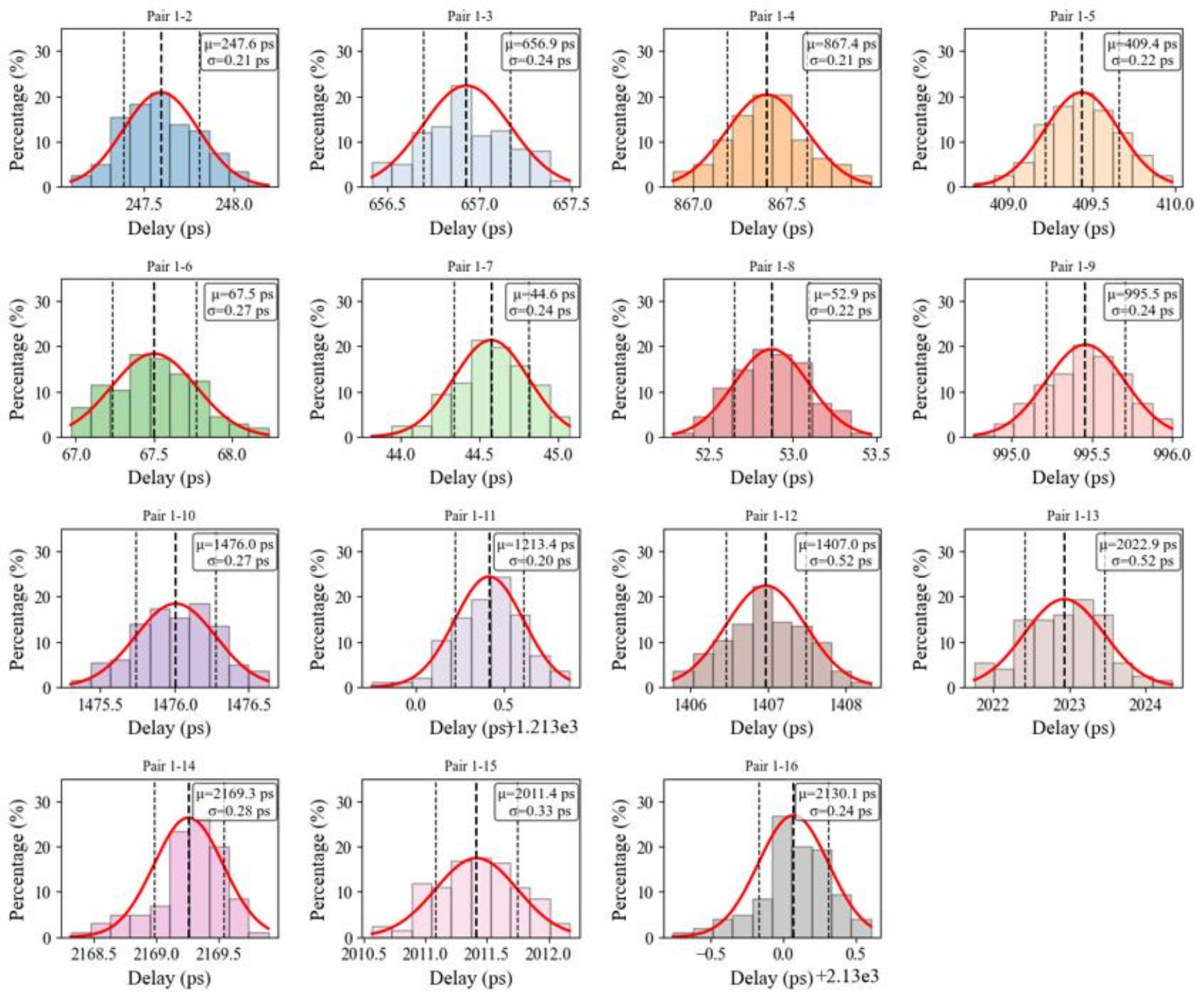}%
\end{center}
\caption{ Histograms of measured delay dispersion between the reference board (Board 1) and each of the remaining fifteen boards (Boards 2–16) within a fully populated crate. Each panel represents the statistical distribution of relative delay extracted from repeated synchronization measurements.}
\label{fig:backplan8}
\end{figure}

\noindent Fig. \ref{fig:backplan8} presents the short-term synchronization analysis between boards within a single crate. The test consisted of a continuous 7-minute acquisition, during which the relative delay between the reference board and each of the other fifteen boards was computed.

\noindent The resulting histograms quantify the temporal dispersion of these delay estimates, characterizing short-term phase stability across the backplane. The narrow, centered distributions observed for all board pairs indicate that the system maintains stable and repeatable latency with minimal jitter (200-500 fs) over minute-scale timescales.

\begin{figure}[H]
\begin{center}
\includegraphics[width=16cm]{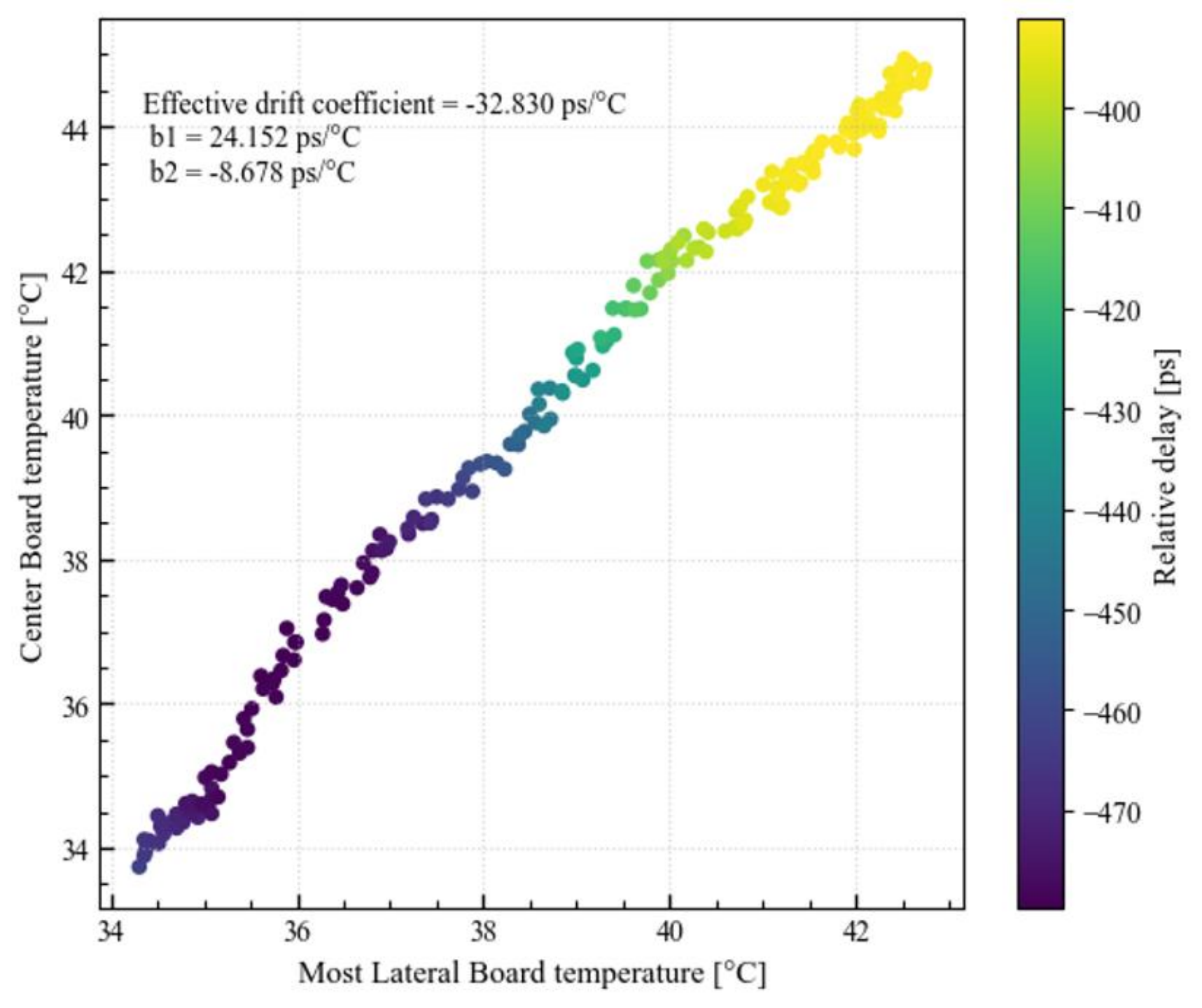}%
\end{center}
\caption{ Relative delay between two CRS boards as a function of their individual board temperatures. The x-axis corresponds to the temperature of the most lateral board, while the y-axis represents the temperature of the center-positioned board. The colormap indicates the measured inter-board delay, highlighting the combined thermal dependence and the effect of differential heating.}
\label{fig:backplan9}
\end{figure}

\noindent The thermal-drift characterization was conducted under a worst-case condition, corresponding to the largest differential temperature within the backplane between a lateral and a central CRS board. Temperature was monitored at the hottest point of the system during correlator firmware operation, defined as the FPGA die temperature. The test ran continuously for 1 hour, during which boards were operated under typical backplane conditions but submitted under distinct thermal profiles due to different positions in the backplane. The results are shown in Fig. \ref{fig:backplan9}.

\noindent A multivariate linear regression $Delay_{ij}=a+b_iT_i+b_jT_j$ separates the contributions of both boards, which $b_i$ and $b_j$, in ps/°C yielded an effective differential drift $b_{eff}=b_j-b_i$ representing the delay change per Celsius degree. Concerning long-term drift, it is important to note that backplane systems are typically operated in thermally regulated environments. Under controlled conditions the residual drift measured here is expected to decrease further, ensuring stable phase alignment over extended operational periods.
However, the broader purpose of this test is to evaluate the differential thermal sensitivity of the backplane–CRS interface. In large interferometric systems, temperature-induced propagation-delay changes are among the dominant non-idealities affecting long-term phase stability. Measuring the relative delay between two boards, rather than the absolute delay of a single path, provides a more relevant metric. The results in Fig.\ref{fig:backplan9} show the expected monotonic dependence of delay on temperature, demonstrating that the timing drift is thermally driven and predictable. This characterization provides a quantitative basis for monitoring, compensating, or mitigating thermal drift in deployed systems.

 \subsection{Backplane System Reliability: Power Delivery and Thermal Performance}
To evaluate the thermal stability and operating margins of the fully populated backplane system under worst-case load, tests were conducted with a fully populated crate (16 CRS boards) operating at full load, exercising high-utilization firmware. Each CRS board employs its own localized cooling solution (bottom-to-top airflow) derived from the CRS thermal management system \cite{Montgomery2024} while the crate fan-tray enhances directed bottom-to-top airflow across the backplane and power modules.
\noindent The system was instrumented using built-in temperature monitoring capabilities,  logging in real time both on-die FPGA junction sensors reading and on-board thermistor measurements over extended runs (duration = 24h, sample period = 60s). The measurements monitored were (i) maximum FPGA temperature ($T_{Jmax}$), (ii) inter-board thermal gradient across slots, and (iii) backplane hotspot temperature near the high-current planes. Experimental results show stable steady-state operation with $T_{Jmax}$ = 52~°C at ambient 21°C, slot-to-slot gradient 14~°C (min=38~°C, max=52~°C), and thermal time constant to 95\% steady state 120~min; crate-level airflow kept power supplies and backplane components well within margin (backplane hotspot 27~°C, $\Delta T_{crate}$ 2.5~°C).
\noindent The crate is powered by a 1.2 kW modular supply (Mean Well NMP1K2 platform). On the power side, we monitored total crate input power, rail regulation, ripple, and drift under 128-input correlator load using commercial lab telemetry. With 16 CRS at full power, the crate draw was 1152 W at 127 VAC (around 72 W per CRS board), with 12 V bus droop 200 mV at load consumption of 92 A and ripple 50m Vpp; per-slot regulators maintained output within 1.5 \% over all test. No protective derating or throttling events were observed, with monitored temperatures consistently below their specified protection thresholds. These results show stable behavior over 24 h of continuous operation and are consistent with CHORD’s extended observational duty cycles.

\section{Conclusion}
This work presents a custom backplane architecture that couples controlled clock distribution with an on-backplane data link operating at 25 Gb/s per lane in a modular, cost-optimized form factor. Designed to address the scaling and timing constraints of large-N radio interferometers, the architecture integrates passive high-speed interconnect, fixed-topology data exchange, and coordinated timing distribution within a compact 6U crate configuration.
\noindent Experimental validation demonstrates stable multi-gigabit link performance, two-digit-picosecond inter-board delay alignment, and jitter on the order of a few hundred femtoseconds, together with comfortable thermal and power margins under full-load operation. These results confirm compatibility with the phase precision required for CHORD-class instruments. Taken collectively, the measurements demonstrate that a requirement-driven passive backplane architecture is sufficient to provide stable, high-bandwidth intra-crate connectivity for scalable correlator systems while maintaining manageable integration complexity.
\noindent The resulting architecture generalizes naturally to other instrumentation requiring synchronized, high-throughput data movement, including TES/MKID spectrometers, passive radiometry chains, and wide-band RFI monitors, by reusing the stable timing network and high-speed data-exchange fabric. To date, the backplane data mesh has been successfully exercised in an 8-board full n² correlator configuration, validating the functionality of the intra-crate data distribution under realistic processing loads. Next steps include extended environmental burn-in validation; initial end-to-end observations are planned for the end of 2026, after which field performance and lessons learned will be reported in a paper documenting the CHORD telescope performance.

\subsection*{Disclosures}
Joshua Montgomery, Graeme Schmecher, and Matt Dobbs are co-founders of t0.technology, the company that developed the backplane technology described in this work. Wellington Avelino received Mitacs support to advance astrophysics applications of these technologies. The authors declare no other conflicts of interest.

\subsection* {Code, Data, and Materials Availability} 
The data supporting the findings of this study are available from the corresponding author upon reasonable request.

\subsection* {Acknowledgments}
The authors gratefully acknowledge funding from Mitacs, the Natural Sciences and Engineering Research Council of Canada, and the Canada Research Chairs program for this research. We also thank CMC Microsystems for granting access to essential software tools used in the simulation and analysis stages of this work.


\bibliography{report}   
\bibliographystyle{spiejour}   


\vspace{2ex}\noindent\textbf{Wellington Avelino} is a researcher and engineer working at the intersection of RF/microwave instrumentation, measurement automation, and control software for advanced sensing systems. He is an author of SPIE Proceedings articles. His current research interests include cryogenic detector readout/control, RF/microwave test and characterization, and low-noise mixed-signal electronics for sensing platforms.

\vspace{1ex}


\end{spacing}
\end{document}